\documentclass[conference]{IEEEtran}
\IEEEoverridecommandlockouts
\usepackage{cite}
\usepackage{amsmath,amssymb,amsfonts}
\usepackage{graphicx}
\usepackage{textcomp}
\usepackage{xcolor}
\usepackage{algpseudocode}
\usepackage{algorithm}
\usepackage{multirow}
\usepackage{float}
\usepackage{subcaption}

\algnewcommand\algorithmicforeach{\textbf{for each}}
\algdef{S}[FOR]{ForEach}[1]{\algorithmicforeach\ #1\ \algorithmicdo}
\def\BibTeX{{\rm B\kern-.05em{\sc i\kern-.025em b}\kern-.08em
    T\kern-.1667em\lower.7ex\hbox{E}\kern-.125emX}}
\begin{document}

\title{Data-driven AC Optimal Power Flow with Physics-informed Learning and Calibrations\\}

\author{\IEEEauthorblockN{1\textsuperscript{st} Junfei Wang}
\IEEEauthorblockA{\textit{Department of EECS} \\
\textit{York University}\\
Toronto, Canada \\
junfeiw@yorku.ca}
\and
\IEEEauthorblockN{2\textsuperscript{nd} Pirathayini Srikantha}
\IEEEauthorblockA{\textit{Department of EECS} \\
\textit{York University}\\
Toronto, Canada \\
psrikan@yorku.ca}
}

\maketitle

\begin{abstract}
The modern power grid is witnessing a shift in operations from traditional control methods to more advanced operational mechanisms. Due to the nonconvex nature of the Alternating Current Optimal Power Flow (ACOPF) problem and the need for operations with better granularity in the modern smart grid, system operators require a more efficient and reliable ACOPF solver. While data-driven ACOPF methods excel in directly inferring the optimal solution based on power grid demand, achieving both feasibility and optimality remains a challenge due to the NP-hardness of the problem. In this paper, we propose a physics-informed machine learning model and a feasibility calibration algorithm to produce solutions for the ACOPF problem. Notably, the machine learning model produces solutions with a 0.5\% and 1.4\% optimality gap for IEEE bus 14 and 118 grids, respectively. The feasibility correction algorithm converges for all test scenarios on bus 14 and achieves a 92.2\% convergence rate on bus 118.
\end{abstract}


\section{Introduction}\label{sec:intro}
With the growing concerns of climate change, the increasing penetration of renewable energy resources, the presence of unpredictable loads, and the proliferation of electric vehicles, the modern electric grid is undergoing a remarkable transition. In power grid operations, Independent System Operators (ISOs) must solve AC Optimal Power Flow (ACOPF) to schedule and control power generation, ensuring that it meets the demand across the entire power grid every minute. ACOPF is one of the significant operation problems solved by ISO to meet electricity demand by scheduling sufficient generations with the minimum cost, which concerns more than 10 billion dollars per year in the U.S. alone\cite{Cain2012}. However, efficiently solving the ACOPF problem has remained a longstanding challenge in power engineering, due to its nonconvex nature, which has been proved to be NP-hard\cite{Lavaei2011}. Moreover, the arise of the new challenges in modern grid enforces the need of fast and reliable ACOPF solvers.

With the development of computational hardware such as Graphics Processing Units (GPUs) and Tensor Processing Units (TPUs), deep learning-based ACOPF has become an area of research interest in recent years. Reference \cite{Zamzam2020} and \cite{Pan2021} proposed using neural networks to predict set-points of generator buses and then obtain full solutions of ACOPF by running traditional power flow solvers (e.g., Newton-Raphson solver) with these set-points as the input to the solver. A random forest-based learning algorithm is proposed in \cite{Kyri2019} to learn warm-start point, which can be used as initial value of traditional ACOPF solvers. However, incorporating traditional ACOPF or power flow solvers in the loop of the problem has concerns on the speed of its convergence, e.g., the speedup of method with traditional solvers is less than one order of magnitude\cite{Huang2021}. 

Some research works target tackling this issue innovatively. A learning algorithm is proposed in \cite{Kyri2022} to imitate the evolution of variables in traditional ACOPF solvers(e.g. Matpower Interior Point Method \cite{Matpower}). This method bypasses the construction process of Jacobian matrices and the inversions. To completely avoid the use of traditional solvers, machine learning models are proposed to predict voltage magnitudes and phase angles in \cite{Huang2021, Rahman2021}. Subsequently, power injections on both generator buses and load buses are recovered based on power flow equations. Nevertheless, inevitable small errors present in voltage prediction by machine learning models result in mismatches when recovering active and reactive power demand at load buses. Similar issue exist in \cite{Pan2023}, where multiple possible voltages as solutions are predicted. Recently, more research methods are proposed to produce full solutions from machine learning models in the literature. A three-stage machine learning scheme is designed in \cite{Lei2020} to decompose the model features for the purpose of reducing the learning difficulties and predicts ACOPF solutions sequentially in different stages. Although it achieves high prediction accuracy for each single variable, the feasibility is not established. Reference \cite{Nellikkath2022,wang2022,Fioretto2020} use the power flow equation as physics-informed soft penalty in the training phase to encourage neural networks learning the physical mappings more effectively. Similarly, the obedience of hard constraints is still an open question for these data-driven ACOPF methods.

In this study, we introduce a physics-informed deep learning framework aimed at generating the optimal solution for the ACOPF problem while enhancing its feasibility. Our approach entails predicting the optimal voltage magnitudes and phase angles at all buses within the target power system, with the learning goal of minimizing both prediction error and physics-informed power injection reconstruction loss. Subsequently, any feasibility-related errors are systematically eliminated using a novel iterative calibration algorithm. This method utilizes Gauss-Seidel updates with warm-start points on load buses while directly adjusting power injection on generator buses. It converges for all test scenarios on the bus-14 grid and achieves a 92.2\% convergence rate on the bus-118 system. These results surpass all state-of-the-art data-driven ACOPF algorithms.

The remainder of this paper is organized as follows. Sec~\ref{sec:form} mathematically defines all notations used in this paper and formulates the ACOPF problem. Then in Sec~\ref{sec:method}, the framework with physics-informed learning and feasibility calibration is proposed. Subsequently, the experimental settings and the efficacy of the proposed algorithm is demonstrated in Sec~\ref{sec:ex}. Finally, we summarize the main insight and illustrate the future extention of our research in Sec~\ref{sec:conc}.  
\vspace{5pt}
\section{Problem Formulation}\label{sec:form}
The fundamental elements in power flow analysis are buses and electrical lines. Buses represent important nodes within the grid, such as generation points, load points, and substations. Each node $i$ is associated with four essential quantities: active and reactive power injections—where on load buses, these quantities are active and reactive power demand ($P_{D_{i}}$ and $Q_{D_{i}}$), while on generator buses, \textcolor{black}{they denote net power injection, which are the differences between active power generation $P_{G_{i}}$ and active power demand $P_{D_{i}}$, and reactive power generation $Q_{G_{i}}$ and reactive power demand $Q_{D_{i}}$}—as well as voltage magnitude ($|V_{i}|$) and phase angle ($\phi_{i}$). The bus set is denoted as $\mathcal{N}=\mathcal{L}\cup \mathcal{G}$, in which $\mathcal{L}$ contains all load buses and $\mathcal{G}$ is the set of all generator buses. The electrical lines and their attributes can be captured by the Nodal Admittance Matrix, which is an $n$-by-$n$ symmetrical complex matrix commonly referred to as the Y-bus matrix. Here, $n=\mathcal{C}(\mathcal{N})$ represents the cardinality of the entire bus set $\mathcal{N}$. In this matrix, the off-diagonal terms $Y_{ij}$ represent the negative value of the line admittance between bus $i$ and bus $j$, equating to zero if no line connects these two nodes. On the other hand, the diagonal term $Y_{ii}$ signifies the summation of the admittances of all lines connected to bus $i$, in addition to the shunt admittance at bus $i$. The input to the ACOPF problem consists of the active and reactive power demand at each bus, i.e., $P_{D_{i}}$ and $Q_{D_{i}}$ for all $i\in \mathcal{N}$. The goal of the ACOPF is to determine a feasible solution that satisfies the given input and all physical constraints, while simultaneously minimizing the generation cost.

The ACOPF problem can be formulated in $\mathcal{P}_{ACOPF}$. For each generator $g\in \mathcal{G}$, the function $C_{g}(\cdot)$ is the cost associated with power generation, which is typically a convex quadratic function of the active power generation $P_{G_g}$. Constraints [C1] and [C2] impose limits on the capacity for active and reactive power dispatch of the generators, respectively. The power flow equation, represented by constraint [C3], ensures the balance of power flow at each single bus in the grid. Constraints [C4] and [C5] govern the inequality constraints for voltage magnitudes and phase angles. Maximum power flow limits for each line is defined by [C6], serving as constraints to prevent overloading.

{\footnotesize
	\begin{align*}
		\mathcal{P}_{ACOPF}:  & \min_{P_{G_{g}}} \sum_{g\in\mathcal{G}}C_{g}(P_{G_{g}}) & \\
		\text{s.t.} \  \forall  \ g \in \mathcal{G}: \ &P_{G_{g}}^{min}\leq P_{G_{g}} \leq P_{G_{g}}^{max} & \textbf{[C1]} \\
		&Q_{G_{g}}^{min}\leq Q_{G_{g}} \leq Q_{G_{g}}^{max} & \textbf{[C2]} \\
		\text{s.t.} \  \forall  \ i \in \mathcal{N}: \ & P_{G_{i}}-P_{D_{i}}+j(Q_{G_{i}}-Q_{D_{i}}) =V_{i}\sum_{j\in \mathcal{N}}V_{j}^{*}Y_{ij}^* & \textbf{[C3]}\\
		&V_i^{min}\leq |V_{i}| \leq V_i^{max} & \textbf{[C4]} \\
		&\phi_i^{min}\leq \phi_{i} \leq \phi_i^{max} & \textbf{[C5]}\\
		&{|V_{i}(V_{i}^{*}-V_{j}^{*})Y_{ij}^*|\leq S_{i,j}^{max} \ \forall \  j \in \mathcal{N},j\ne i }& \textbf{[C6]}
	\end{align*}
}
Additionally, the constraints [C3], [C4] and [C6] collectively form non-convex constraints in ACOPF problem, make it a NP-hard problem in the worst-case\cite{Lavaei2011}. \\

\vspace{5pt}
\section{Proposed Algorithm}\label{sec:method}
In this section, we present a comprehensive two-stage framework designed to efficiently generate solutions to the ACOPF problem. The first stage focuses on developing an algorithm to learn mappings between power demand and the optimal voltage at all buses within the grid. Additionally, it determines the optimal power dispatch from each generator. Subsequently, the second stage utilizes the power flow equations to refine and calibrate the outcomes produced in the first stage. Finally, the resulting solution from these two stages establishes feasibility with a high success rate and a negligible gap from the optimal solution.

\subsection{Stage 1: Physics-informed Deep Learning for ACOPF Prediction}
In this stage of our research, we leverage the power of deep neural networks (DNNs) to capture the underlying relationship between power demand ($P_{D_{i}}$ and $Q_{D_{i}}$) and the optimal voltage magnitude ($|V_{i}|$) and phase angles ($\phi_{i}$) at each bus $i \in \mathcal{N}$ within the grid. These voltage magnitudes and phase angles constitute only a portion of the decision variables. Subsequently, the complete solution is derived by running power flow equations, encompassing the active/reactive power dispatch ($P_{G_{g}}$ and $Q_{G_{g}}$) at all generator buses $g\in\mathcal{G}$. As introduced in Section~\ref{sec:intro}, data-driven models commonly treat ACOPF as a black-box function, taking input as the power demand and generating optimal solutions as output. However, it is vital that these solutions adhere to the all constraints [C1]-[C6] in $\mathcal{P}_{ACOPF}$. Previous approaches, such as that proposed by in \cite{Pan2021}, separate the ACOPF variables into independent variables predicted by DNNs and dependent variables calculated through conventional iterative power flow solvers. Yet, these solvers experience slow convergence, e.g., they only accelerate the ACOPF by less than one order of magnitude\cite{Huang2021}, especially when applied to large power grids. Hence, generating complete ACOPF solutions without traditional solvers is preferred. DeepOPF-V \cite{Pan2023} aims to produce optimal voltages from DNN models, but the side effect is the unavoidable errors on power demand at load buses.

\begin{figure}[tb]
	\centerline{\includegraphics[scale=0.3]{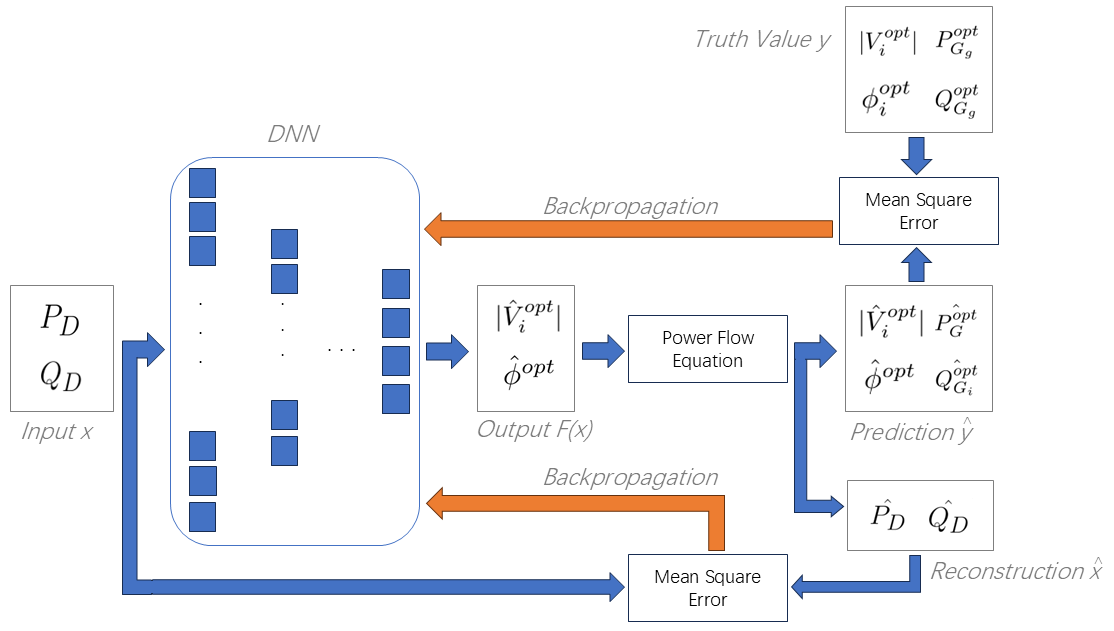}}
	\caption{Diagram of End-to-end Learning for ACOPF.}
	\label{fig:end2end}
\end{figure}

We present an end-to-end DNN model as illustrated in Fig. \ref{fig:end2end}. The model is designed to first forecast optimal voltage magnitudes and phase angles given a specific power demand as input. Subsequently, it runs power flow equations to calculate power generation and reconstruct the power demand simultaneously. The error on each variable is then measured as prediction error and physics-informed penalty within the learning cost function. 
The input $x:\{P_{D_{i}},Q_{D_{i}},i\in\mathcal{N}\}$, encompassing active and reactive power demand for all buses, undergoes processing by a DNN denoted as $F(\cdot)$. The output $F(x)$ is the optimal voltage magnitudes and phase angles, i.e., $F(x):\{|\hat{V_{i}}^{opt}|,\hat{\phi_{i}}^{opt},i\in\mathcal{N}\}$, which is concatenated with the optimal active and reactive power dispatch on each generator bus calculated by power flow equations defined in Eq.~\ref{eq:pg} and \ref{eq:qg}, to form the full solution $\hat{y}:\{|\hat{V}_{i}^{opt}|,\hat{\phi}_{i}^{opt},\hat{P}_{G_{g}}^{opt},\hat{Q}_{G_{g}}^{opt},i\in \mathcal{N},g\in\mathcal{G}\}$ of ACOPF. 
\begin{align}\label{eq:pg}
	&\hat{P}_{G_{g}}^{opt} = P_{D_{g}} + Re(\hat{V}_{g}^{opt}\sum_{j\in\mathcal{N}} \hat{V}_{j}^{opt^*}Y_{gj}^* )\\
	&\hat{Q}_{G_{g}}^{opt} = Q_{D_{g}} + Im(\hat{V}_{g}^{opt}\sum_{j\in\mathcal{N}} \hat{V}_{j}^{opt^*}Y_{gj}^* )\label{eq:qg}
\end{align}
Meanwhile, the inputs of the ACOPF problem, i.e. active and reactive power demand on all buses are reconstructed as $\hat{x}:\{\hat{P}_{D_{i}},\hat{Q}_{D_{i}},i\in \mathcal{N}\}$ by Eq.~\ref{eq:pd_recon} and \ref{eq:pd_recon1}. 
\begin{align}\label{eq:pd_recon}
	&\hat{P}_{D_{i}} = \hat{P}_{G_{i}}^{opt}-Re(\hat{V}_{i}^{opt}\sum_{j\in\mathcal{N}} \hat{V}_{j}^{opt^*}Y_{ij}^* )\\
	&\hat{Q}_{D_{i}} = \hat{Q}_{G_{i}}^{opt}-Im(\hat{V}_{i}^{opt}\sum_{j\in\mathcal{N}} \hat{V}_{j}^{opt^*}Y_{ij}^*)\label{eq:pd_recon1}
\end{align}

The proposed mechanism is trained to ensure that the prediction outputs $\hat{y}:\{|\hat{V}_{i}^{opt}|,\hat{\phi}_{i}^{opt},\hat{P}_{G_{g}}^{opt},\hat{Q}_{G_{g}}^{opt},i\in \mathcal{N},g\in\mathcal{G}\}$ closely approximate the outputs from commercial ACOPF solvers $y:\{|V_{i}^{opt}|,\phi_{i}^{opt},P_{G_{g}}^{opt}, Q_{G_{g}}^{opt},i\in \mathcal{N},g\in\mathcal{G}\}$. Additionally, it minimizes the error between the input variables $x:\{P_{D_{i}},Q_{D_{i}},i\in\mathcal{N}\}$ and the reconstructed inputs $\hat{x}:\{\hat{P}_{D_{i}},\hat{Q}_{D_{i}},i\in\mathcal{N}\}$, thus establishing a cycle-consistency. This cycle-consistency enhances an accurate mapping from the input to the output and back to the input, maintaining coherence throughout the process.

The error between $\{x,y\}$ and $\{\hat{x},\hat{y}\}$ is calculated via mean square error loss, defined in Eq.~\ref{eq:mse}. 
\begin{equation}\label{eq:mse}
	\begin{split}
		L_{mse}=&\mathbb{E}_{i\in \mathcal{N}}[(\hat{P}_{D_{i}}-P_{D_{i}})^{2}+(\hat{Q}_{D_{i}}-Q_{D_{i}})^{2}\\
		&+(|\hat{V}_{i}^{opt}|-|V_{i}^{opt}|)^{2}+(\hat{\phi}_{i}^{opt}-\phi_{i}^{opt})^{2}]\\
		&+\mathbb{E}_{g\in \mathcal{G}}[(\hat{P}_{G_{g}}^{opt}-P_{G_{g}}^{opt})^{2}+(\hat{Q}_{G_{g}}^{opt}-Q_{G_{g}}^{opt})^{2}]\\
	\end{split}
\end{equation}

Then, the learnable weights in the DNN will be updated via back-propagation algorithm\cite{Goodfellow2016}, which calculates the gradient of the overall loss function $L_{mse}$ w.r.t each weight in the model and updates the weights to minimize the learning loss.  

\subsection{Stage 2: Feasibility Calibration Algorithm}
While physics-informed deep learning has shown superior performance compared to purely data-driven approaches\cite{Fioretto2020,Nellikkath2022}, the output of a DNN inevitably contains errors. These errors are common even for training data. Consequently, in the context of the ACOPF problem, these errors cause the input parameters $x$ and decision variables $\hat{y}$ to deviate from strict adherence to the equality constraint [C3]. For example, Huang,et al. \cite{Huang2021} claims that a 1\% imbalance on load buses is acceptable and can only be eliminated by controllable distributed energy sources. In this section, we introduce a feasibility calibration algorithm aimed at perturbing the output from the stage 1 to enhance the adherence to the power flow equations while avoiding violation of other constraints.

For each load bus $i \in \mathcal{L}$, the voltage from the stage 1 can be updated based on the rule defined in Eq.~\ref{eq:fpi}. This is the rule used in Gauss-Seidel algorithm\cite{Book1999}, derived from [C3], leveraging the slack of voltage value on each load bus for a local adjustment based on the current state of the load bus itself and its neighbours.
\begin{equation}\label{eq:fpi}
{\hat{V}_{i,new}}^{opt}=\frac{1}{Y_{ii}}\big(\frac{-P_{D_{i}}+jQ_{D_{i}}}{{{{\hat{V}_{i,old}}^{opt}}}}-\sum_{j\in \mathcal{N}, i\neq j}\hat{V}_j^{opt}Y_{ij}\big)
\end{equation}
The power injection at each generator bus $g\in \mathcal{G}$ is straightforwardly updated to eliminate all imbalances, as defined in Eq.~\ref{eq:pg} and \ref{eq:qg}.

The proposed algorithm operates in epochs. Within each epoch, voltages on all load buses are updated at first, followed by adjustments to power injections at generator buses. The algorithm exits either when a predefined stop criterion $\rho$ for average power flow imbalance at each bus is met or when the maximum epoch $E$ is reached. The algorithmic framework is summarized in Alg.~\ref{alg:alg1}. The $Clip$ function is defined in Eq.~\ref{eq:clip}, where the target variable is $z$, whose upper and lower bounds are $\overline{z}$ and $\underline{z}$ respectively. The $Check$ function returns residuals regarding the constraint [C3]. 
\begin{equation}\label{eq:clip}
	Clip(z)=\max(\min(z,\overline{z}),\underline{z})
\end{equation}

We have following three advantages in this algorithm compared to the Gauss-Seidel algorithm. Firstly, we initiate the calibration process based on the results from the first stage, incorporating complete variables in an ACOPF solution that are already very close to convergence. In contrast, Gauss-Seidel only utilizes two variables at each bus, with the others starting from random guesses. Secondly, our proposed method updates from load buses to generator buses, allowing imbalances to be absorbed by the generator buses. Finally, instead of fixing the active power injection at generator buses, we adjust both active and reactive power injections to utilize the slack on generator buses, leading to faster convergence.

\begin{algorithm}[htp!]
	\footnotesize
	\begin{algorithmic}[1]
		\caption{Calibration of solution for ACOPF.}
		\label{alg:alg1}
		\Function{correction}{$x,\hat{y},E,\rho$}
		\State $e\leftarrow 0$
		\State r = $Check$ [C3]
		\While{ $|r|>\rho$ \textbf{ and } $e< E$}
		\State $e\leftarrow e+1$
		\ForEach{ $i \in$ $\mathcal{L}$}
		\State ${\hat{V}_{i,new}}^{opt}=\frac{1}{Y_{ii}}\big(\frac{-P_{D_{i}}+jQ_{D_{i}}}{{{{\hat{V}_{i,old}}^{opt}}}}-\sum_{j\in \mathcal{N}, i\neq j}\hat{V}_j^{opt}Y_{ij}\big)$
		\State $Clip$(${\hat{V}_{i,new}}^{opt}$)
		\EndFor
		\ForEach{g $\in$ $\mathcal{G}$}
		\State $\hat{P}_{G_{g}}^{opt} = P_{D_{g}} + Re(\hat{V}_{g}^{opt}\sum_{j\in\mathcal{N}} \hat{V}_{j}^{opt^*}Y_{gj}^*)$\\
		\State  $\hat{Q}_{G_{g}}^{opt} = Q_{D_{g}} + Im(\hat{V}_{g}^{opt}\sum_{j\in\mathcal{N}} \hat{V}_{j}^{opt^*}Y_{gj}^*)$
		\State Clip($\hat{P}_{G_{g}}^{opt})$, Clip($\hat{Q}_{G_{g}}^{opt}$)
		\EndFor
		\State r = $Check$ [C3]
		\EndWhile	
		
		\EndFunction
	\end{algorithmic}
\end{algorithm}

\section{Experiment}\label{sec:ex}

In this section, we will first detail the environment used in this research, including the software, libraries, computation platform, etc. Then, we will illustrate the numerical results of the proposed method.

\subsection{Environment and Dataset}
In this work, the data generation process was executed on a local laptop with an Intel Core i7-10750H CPU and 32.0 GB RAM. The algorithm was implemented in Python with Tensorflow and Keras 2.7.0. The training process itself was conducted on Google Colaboratory, a cloud-based platform offering access to Google's formidable computing resources including GPUs and TPUs. In this work, NVIDIA Tesla V100-SXM2-16GB is used. To assess the performance of our proposed algorithm, we conducted all experiments on two standard benchmark systems: IEEE bus-14 and IEEE bus-118.

The dataset utilized in this study, comprising ACOPF solutions, was generated through the utilization of MATPOWER's primal-dual interior point solver (MIPS)\cite{Zimmerman2010}. We introduced variability by randomly perturbing the nodal active and reactive power demands. This perturbation ranged from 80\% to 120\% of their respective nominal values, thereby simulating real-world fluctuations in power demand. This is also the common practice in the literature,e.g. in \cite{Fioretto2020,Pan2021,Pan2023}. For each of these benchmark systems, we generated a dataset comprising 100K pairs of input and optimal output data. This extensive dataset was then divided into two subsets: 80\% for training purposes and 20\% for testing.

\subsection{Performance of the Proposed DNN} 
The proposed DNN model is selected to be a Convolutional Neural Network(CNN), which uses convolutional filters to extract meaningful patterns and features from input data. Comparing to fully-connected neural networks, CNNs are more weight-efficient due to its design of weight sharing and sparse connectivity\cite{Goodfellow2016}. To remove the inequality constraints on individual decision variables defined on [C4] and [C5], we introduce sigmoid activation functions in the output layer of the DNN. The sigmoid function's S-shaped curve, bounded between 0 and 1, allows us to map the DNN's output into the permissible range specified by the lower and upper bounds of the variables. This mapping is captured by Eq. \ref{eq:last_layer}, where $F_{k}^{L}$ denotes the k-th output from the final layer through a sigmoid function, $x_{k}^{min}$ and $x_{k}^{max}$ are lower and upper bounds for each variable.

\begin{equation}\label{eq:last_layer}
	F_{k}(x)={F_{k}^{L}}\cdot (x_{k}^{max}-x_{k}^{min})+x_{k}^{min}
\end{equation}
The model with different hyperparameters including number of layers, number of filters, activation function and batch size are used to validate the performance so as to select the best model. The settings of these models are listed in Tab.~\ref{tb:settings}. The model selection is based on their performance on the validation set. The model with the best performance for each testing grid are labelled with $\dagger$, and selected to be the final model $F(\cdot)$.

\begin{table}[htbp]
	\caption{Different Settings of DNN's Hyperparameter}\label{tb:settings}
	\begin{center}
		\begin{tabular}{|c|c|c|c|c|}
			\hline
			\textbf{Power Grid} & \textbf{Settings} & \textbf{Layers} & \textbf{Activation} & \textbf{Batch}\\
			\hline
			\multirow{4}{*}{\centering 14} & model-1 $\dagger$ & $512|256|128|64$ & LeakyReLU & 512\\
			& model-2 & $256|128|64|32$ & ReLU & 512 \\
			& model-3 & $768|512|128$ & LeakyReLU & 256 \\
			& model-4 & $64|32$ & ReLU & 128 \\
			\hline
			\multirow{4}{*}{\centering 118} & model-1 & $512|256|128|64$ & LeakyReLU & 256\\
			& model-2$\dagger$ & $512|256|128|64$ & LeakyReLU & 512 \\
			& model-3 & $768|512|128$ & LeakyReLU & 256 \\
			& model-4 & $64|32$ & ReLU & 128 \\
			\hline
		\end{tabular}
	\end{center}
\end{table}	

The model is composed of 189,148 parameters for the IEEE bus-14 system and 309,164 for the IEEE bus-118 system. This architectural choice provides expressive power required to capture the relationship of the underlying data effectively.

To provide a comparative study of our proposed approach, we adopt the same DNN architecture to implement benchmark algorithms, such as supervised learning\cite{Guha2019} and DeepOPF-V \cite{Huang2021}. Unlike original approaches that calculate each variable from a separate DNN, we opt for a unified approach, utilizing a single DNN to predict all variables. This streamlined methodology facilitates a direct comparison of methods with similar computational complexity, thereby enabling a fair assessment of performance.

\begin{table*}[tp]
	\vspace{3pt}
	\caption{Comparison of DNN's Performance in Different Algorithms}
	\begin{center}
		\begin{tabular}{|c|c|c|c|c|c|c|c|c|}
			\hline
			\multirow{2}{*}{\textbf{Power Grid}} & \multicolumn{2}{|c|}{\textbf{Architecture}} & \multicolumn{6}{|c|}{\textbf{Performance}} \\
			\cline{2-9} 
			& \textbf{\textit{Model Type}}& \textbf{\textit{\#Parameters}}& \textbf{\textit{$|V|_{mse}$}} & \textbf{\textit{$\phi_{mse}$}} & \textbf{\textit{${P_{G}}_{mse}$}} & \textbf{\textit{${Q_{G}}_{mse}$}} & \textbf{\textit{Feasibility}} &Optimality Gap\\
			\hline
			\multirow{3}{*}{14} 
			& Supervised learning\cite{Guha2019}&190,168 & $5.2e^{-4}$&$3.2e^{-4}$ &$3.5e^{-4}$ & $7.2e^{-5}$&$8.5e^{-3}$&2.25\%\\
			\cline{2-9}
			& DeepOPF-V\cite{Pan2023}&189,148& $6.8e^{-6}$& $6.2e^{-6}$& $8.3e^{-5}$ & $1.9e^{-5}$ & $3.5e^{-3}$&0.72\%\\
			\cline{2-9}
			& Proposed &189,148 & $1.0e^{-4}$& $1.2e^{-4}$& $7.1e^{-5}$ & $3.8e^{-5}$ & $<1e^{-6}$&0.67\%\\
			\hline
			\multirow{4}{*}{118} &Supervised learning\cite{Guha2019}&324,504 & $3.5e^{-4}$&$1.4e^{-3}$ &$5.3e^{-4}$ &$5.5e^{-4}$ & 0.032&0.17\%\\
			\cline{2-9}
			& DeepOPF-V\cite{Pan2023}&309,164& $1.3e^{-4}$& $2.3e^{-4}$& $0.042$ & $3.7e^{-3}$ & $0.039$&1.03\%\\
			\cline{2-9}
			& Proposed& 309,164& $3.8e^{-4}$&$5.8e^{-4}$ & $2.82e^{-3}$& $3.1e^{-4}$& $<1e^{-6}$ &0.15\%\\
			\hline
		\end{tabular}
		\label{tb:comp}
	\end{center}
\end{table*}

\begin{figure*}[htbp]
	\centering
	\begin{subfigure}{0.4\textwidth}
		\centering
		\includegraphics[width=\linewidth]{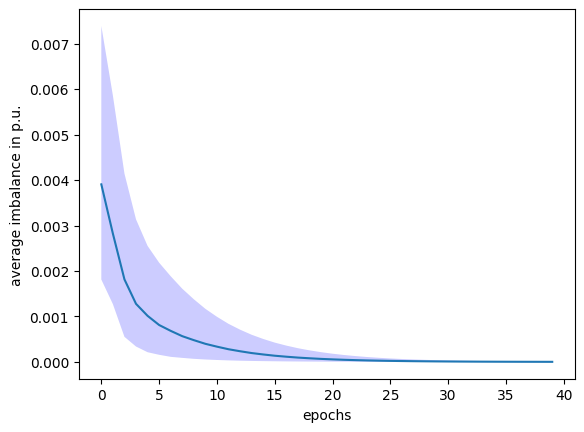}
		\caption{Calibrations on IEEE Bus-14.}
		\label{fig:cali}
	\end{subfigure}
	\begin{subfigure}{0.4\textwidth}
		\centering
		\includegraphics[width=\linewidth]{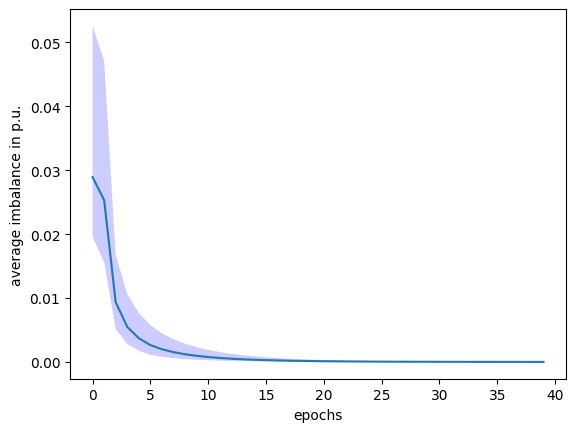}
		\caption{Calibrations on IEEE Bus-118.}
		\label{fig:cali118}
	\end{subfigure}
	\caption{Statistical Calibration Processes}
	\label{fig:calibration}
\end{figure*}
The detailed results of the comparative evaluation conducted on the IEEE bus-14 and IEEE bus-118 systems are presented in Table~\ref{tb:comp}. For the bus-14 system, all three algorithms achieved prediction errors of approximately $1e^{-4}$ for each variable, but our proposed method achieves a feasibility level of $1e^{-6}$ regarding constraint [C3]. When considering the optimality gap, our proposed method outperforms the others. Moving to the bus-118 system, DeepOPF-V tends to overfit voltages, lacking in capturing the underlying physical dynamics necessary for recovering power injection. Supervised learning method encounters challenges in accurately predicting all variables while ensuring feasibility. However, our proposed algorithm demonstrates more stable prediction errors, lower feasibility, and a smaller optimality gap. Therefore, our proposed model emerges as a more robust and effective solution, surpassing both counterparts across various metrics.

\subsection{Performance of the Calibration Algorithm}
In Alg.~\ref{alg:alg1}, we specify the early stopping criterion $\rho$ as $1e^{-6}$, ensuring that the calibration process terminates when the change in the objective function becomes negligible. Additionally, we set the maximum number of epochs for calibration $E$ to 100, providing a balance between computational efficiency and convergence.

The calibration algorithm is evaluated using 10,000 testing points for both the IEEE bus-14 and IEEE bus-118 systems. Remarkably, we achieve a convergence rate of 100\% for the IEEE bus-14 system and 92.2\% for the IEEE bus-118 system, demonstrating the effectiveness of our approach across diverse scenarios. To visualize the convergence behavior, we plot the average performance curve for the convergence cases with a solid line in Fig.~\ref{fig:cali} and \ref{fig:cali118}. The plot illustrates how imbalance in the grid is eliminated across epochs. Additionally, the plot includes a shaded interval between the best and worst cases for each epoch, demonstrating the statistical patterns of the convergence algorithm across different scenarios. Convergence with average imbalance below $\rho$ is typically achieved around the 33rd epoch for the IEEE bus-14 system and the 40th epoch for the IEEE bus-118 system.

Furthermore, we assess the optimality of the calibrated ACOPF solutions to ascertain the degree of perturbation introduced by the calibration process. Remarkably, our proposed algorithm exhibits very small perturbation, with the average optimality gap experiencing only slight changes. For instance, the average optimality gap increases from 0.58\% to 0.67\% for the IEEE bus-14 system and decreases from 0.20\% to 0.15\% for the IEEE bus-118 system. These negligible changes show the ability of our algorithm to refine ACOPF solutions while preserving their optimality.

\section{Conclusion and Future Work}\label{sec:conc}
In conclusion, we proposed a physics-informed DNN framework for addressing ACOPF problem in power grid operations. By leveraging DNN and incorporating physics-informed penalties, we have effectively captured the complex relationships between power demand, optimal voltage magnitudes, and optimal phase angles, while adhere to critical constraints and operational requirements. Moreover, our calibration algorithm has showcased 100\% convergence rates on bus 14 system and 92.2\% on bus 118 system with negligible perturbations to ACOPF solutions from DNN, thus demonstrated its practical utility in real-world scenarios. In future work, we will investigate robust machine learning with uncertainties on input. This is practical problem since power grid operations are mainly based on load forecasting. Secondly, unsupervised learning or semi-supervised learning without the use of traditional ACOPF solvers for labelled data generations would be also an extensive direction of this work. For feasibility studies, deriving theoretical conditions to guarantee the convergence would be a significant contribution to the area of ACOPF.

\vspace{12pt}

\end{document}